\documentclass[11pt]{article}
\usepackage{graphicx}
\begin{document}
\section*{\footnotesize{\it Indian Journal of Theoretical Physics} {\bf 48},
 no. 2, pp. 97-132 (2000)}

\vspace{6mm}

\bigskip\medskip
\begin{center}
{\LARGE \bf On the nature of spin, inertia and gravity
\\ of a moving canonical particle}
\end{center}

\vspace{2mm}
\begin{center}
{\bf Volodymyr Krasnoholovets}
\end{center}

\begin{center}
{Institute of Physics, National Academy of Sciences, \\ Prospect
Nauky 46,   UA-03028 Ky\"{\i}v, Ukraine \ \
http://inerton.cjb.net}
\end{center}

\vspace{4mm}
\begin{center}
\ \ \ \ \ \ \ \ \ \ \ \ \ \ \ \ \ \ \ \ \ \ \ \ \ \ \ \ \ \ \ \ \
\ \ \ \ \ \ \ \ \ \ \ \ \ \ \ \ \ \ \ \ \ \ \ \ \ \ \ \ \ \ \ \ \
\ \ \ \ \ \ \ \  October 1998 -- November 1999
\end{center}

\begin{center}
{\small{\bf Abstract}}
\end{center}
{\small

 It is suggested that a moving canonical particle interacts
with a vacuum regarded as a "soft" cellular space. The interaction
results into the emergence of elementary excitations of space --
inertons -- surrounding the particle. It is assumed that such a
motion leads not only to the spatial oscillation of the particle
along a path but to the oscillation of the particle centre-of-mass
as well. This phenomenon culminating in the anisotropic pulsation
of the particle is associated with the notion of spin. The
particle-space interaction is treated as the origin of the matter
waves which are identified with the particle inertia and inertons
surrounding the moving particle are considered as carriers of its
inert properties. Inertons are also identified with real carriers
of the gravitational interaction and the range of the particle
gravitational potential is evaluated by the inerton cloud
amplitude $\Lambda=\lambda c/v_0$, where $\lambda$ is the de
Broglie wavelength, $c$ and $v_0$ are the velocity of light and
the particle respectively.  The nature of the phase transition
that occurs in a quantum system when one should pass from the
description based on the Schr\"odinger formalism to that of
resting on the Dirac one is explained in detail. }

\vspace{3mm}

\begin{description}
\item [{\bf Key words:}]
quantum mechanics, spin, space, inertons, gravitation \\

\vspace{2mm}

\item[\sf {\bf PACS:}] 03.65.w Quantum mechanics;  \\ 04.50 Unified field
theories and other theories of gravitation;  \\ 04.60.-m Quantum
gravity
\end{description}

\newpage

\section{\bf Introduction and the statement of problem}

\vspace{2mm}\hspace*{\parindent}

    Since the introduction of a hypothesis for the spin by Goudsmit [1] and
Uhlenbeck [2] (see also Van der Waerden [3]), the question as to
the nature of this phenomenon has consistently been a major focus
of interest for researchers. Most of them are guided by the
findings of Schr\"odinger [4] and subsequently of Dirac [5] who
have shown that a rapid oscillatory motion -- {\it Zitterbewegung}
-- is peculiar to a free relativistic particle. Then Frenkel [6]
has assumed that the spin can be regarded as a proper mechanical
moment, which is due to the fuzzy mass and charge hidden from an
observer. The ideas of Schr\"odinger and Frenkel have stimulated
the emergence of such particle spin models as bilocal rotator,
Halbwachs et al. [7], relativistic oscillator, Ginzburg and Man'ko
[8], and stochastic oscillator, Petroni et al. [9], dequantized
spin-particle, Plahte [10], and extended objects, Umezawa [11],
Barut and Tracker [12], Bohm et al. [13]. Barut and Zanghi [14]
have considered the spin as an angular moment of the real {\it
Zitterbewegung}. Spavieri [15] has described the spin of the
particle by the intrinsic component, a massless subparticle that
travels around the centre-of-mass of the particle in a fuzzy
orbit. Berezin and Marinov [16] and Srivastava and Lemos [17] have
treated the spin as a construction of hidden variables that obey
Grassmann's algebra. Kuryshkin and Entralgo [18] have discussed
the classical spin model based on the concept of a
structural-point object that implies an assembly of point
particles with distinctive small masses and charges confined to a
specific interaction, which operates between them. On the other
hand, Ohanian [19] includes the assumption that neither the spin,
nor the magnetic moment of the electron are its intrinsic
properties and that they are nothing but the structure of the
field of the electronic wave. By assuming that the intrinsic
magnetic moment of the electron is the initial notion, Heslot [20]
associates the emergence of the spin with the generator of a group
of rotations in nonrelativistic classical mechanics. Presented by
Shima [21] is the field Lagrangian that satisfies the gauge
invariance and describes the spin-1/2 massless Dirac particle.
Pav$\rm \check s$i$\rm \check c$ {\it et al}. [22] have developed
the model proposed by Barut and Zanghi [14] in terms of Clifford
algebra; Pav$\rm \check s$i$\rm \check c$ et al. have obtained a
non-linear Dirac-like equation which characterises a cylindrical
helix trajectory of a point-like object, i.e. spinning particle.
Corben [23] has proposed an interesting approach to the structure
of a spinning point particle at rest. Sidharth [24] has
conjectured that spin is a consequence of a space-time cut off at
the Compton wavelength and Compton time scale. Oudet [25] has
maintained the deep analysis of experimental manifestations of the
spin and the model of the electron in which the electron is
thought as a small fluid mass has been discussed.  The shape of
the mass is supposed to be transformed according to the energy
exchanged between the field and the electron along its trajectory.
This conception proposes that there is energy associated with the
spin. It is notable that the author's following study is roughly
similar to the pattern described by Oudet.

      In the above papers, a variety of notions, in some instances mutually
exclusive ones, provide the basis for the development of the spin model,
and yet some specificity that must necessarily be inherent in an elementary
particle is the initial point. The present paper also deals with the spin
problem. However the author pursues an object to study the problem in the
framework of a vacuum medium model. And in this case the so-called hidden
variables (see, e.g. Bohm [26]) evidently should play a primordial
important role.

     There are several new views on the vacuum substance (see, e.g.
Fomin [27], Aspden [28], and Vegt [29]).  On special note is approaches
developed by Winterberg [30] and Rothwarf [31].
Winterberg's aether is a densely filled substance with an equal number
of positive and negative Planck masses $m_{\rm P}=\sqrt{hc/G}$ which
interact locally through contact-type delta-function potentials; in the
framework of this approach Winterberg has shown that quantum mechanics
can be derived as an approximate solution of the Boltzmann equation for
the Planck aether masses. The particle in that model is a formation
appeared owing to the interaction between the positive and negative
Planck masses similar to the phonon in a solid.

The Rothwarf's aether model [31] is based upon a degenerate
Fermion fluid, composed primarily of electrons and positrons in a
negative energy state relative to the null state or true vacuum. A
key assumption of Rothwart was that the speed of light was the
Fermi velocity of the degenerate electron-positron plasma that
dominates the aether. The model was applied for the description of
a large number of phenomena such as the nature of spin (considered
as a vortex in the aether), electric fields (polarization of the
aether), the nature of the photon (a region of rotating polarized
aether propagating with a screw-like motion), etc.

     The author's vacuum concept [32-34] is close to that of Winterberg
[30].  However, unlike Winterberg's aether, it supposes an
elasticity of discrete vacuum substance and takes into account the
possibility of deterministic motion of a particle in real space.
It is the first physical model of space that is not contradictory
to the formalism of special relativity (see Ref. [33]). At the
same time the concept conjectures a new type of elementary
excitations of space -- inertons, -- which should substitute for
gravitons of general relativity. The statement needs an
explanation. As we know classical mechanics was the origin both of
the theory of relativity and quantum mechanics. In classical
mechanics an object's mass can manifest itself as gravitational
mass (in gravitational experiments) or as inert one (in the
dynamics including the kinetics). General relativity asserts (see,
e.g. Bergmann [35]) that gravitational and inert mass are
equivalent. However, the affirmation was proved in the case of
macroscopic phenomena: The theory of relativity derives the
equations of motion of a massive point in the highly rarefied
space (see, e.g. Bergmann [36], Pauli [37], Weinberg [38],
Dubrovin et al. [39]). Recent developments in quantum general
relativity (see, e.g. Ashtekar [40]) do not concern this
significant problem as well.

    Quantum mechanics does not tell one notion of mass from the other
one and in addition the equations of quantum mechanics  include
mass only as a pure classical parameter. So orthodox quantum
theory can say nothing about the mass behaviour in atom and
nucleus range. Sidharth [41] tried to study this problem; he
concluded that inert mass of an elementary particle is the energy
of binding of nonlocal amplitudes in the zitterbewegung Compton
wavelength region. But what energy is?! When applied to a
particle, the energy means the square form: the mass times the
velocity to the second power. Much probably in this case just the
notion of mass is initial. What is more, the submicroscopic
construction of quantum mechanics developed at the scale of
$10^{-28}$ cm in the author's research [32-34] evidences that the
notion of inert mass is original.  The moving particle interacts
with space which is not empty but fine-grained, in any event
"so-called empty space is actually filled with elementary
particles that pop in and pop out and ...  these "virtual"
particles ... could exert a gravitational force", Krauss [42]). As
a consequence, elementary deformation excitations caused by that
interaction must accompany the moving massive point.  This means
that the Einstein classical equations which describe the mass
distribution in empty space should not only be added by the lambda
term but they are bound to be change:  The equations should be
constructed with due regard for the interaction of the moving
mass/masses with space as well.  When such is the case, the
mentioned excitations, i.e. inertons, will play a role of actual
carriers of gravitational interaction. It must be emphasized that
clouds of inertons surrounding electrons, indeed, manifest
themselves experimentally; the statement has recently been proved
by the author[43].

     The author has previously demonstrated [32,33] that the mechanics
of a massive point, which was interacting with space, culminated in
the Schr\"odinger wave equation formalism. In the present paper the
investigation of the behaviour of a massive particle with the spin is
performed and it is shown how such kind of the submicroscopic mechanics
can give rise to the Dirac equation formalism. Besides the proposed
paper gives a coherent explanation, as compared to author's preceding
papers on the inerton motion, notions of space, and the space
crystallite and peculiar features of the latter are discussed.

\vspace{2mm}

\section{\bf Intrinsic motion of a particle}

\vspace{2mm} \hspace*{\parindent}
     Space is treated by the author [32-35] as a specific medium (a
quantum aether) with a cellular structure, each cell being either
the superparticle or corpuscle filled (i.e. this is a mathematical
space constructed of balls). The corpuscle is thought to evolve
from the superparticle when degeneration over one of its multiplet
states is eliminated. The velocity of light $c$, i.e. the speed of
the transmission of information from superparticle to
superparticle, is a property of space. Superparticles do not
transfer in space and yet they are capable of conveying
information by a relay mechanism. As for the corpuscle, it
actually travels passing between fluctuating superparticles.  We
will identify the corpuscle with a canonical particle like the
electron, muon, and so on.

     In the present model, the initial velocity $v_0$ of the particle is
not constant along its trajectory due to permanent mutual
collisions with superparticles. If the particle is not absolutely
rigid it can be presumed that along with a nonstationary
translational movement, the particle also exhibits a nonstationary
intrinsic motion, i.e. owing to mutual collisions with space,
radial pulsation is excited in the particle (a relative motion of
the front and back surfaces along the line coincident with the
particle trajectory).

     In paper [33] the Lagrangian of a relativistic spinless
particle has been chosen in the form:
\begin{equation}
 L=-M_0 c^2 \Bigl\{ 1- \bigl[ \dot X^2+\sum_r U(X, \dot X; x_{(r)},
 \dot x_(r) ) \bigr] /c^2 \Bigr\}^{1/2}
\label{1}
\end{equation}
where $M_0$ is the mass of the particle at rest. The function $U$
allows for the kinetic energy of the particle-emitted inerton
ensemble and the potential of the interaction between the particle
and inertons. In (1) the translational motion of the physical
"point" (particle cell) along its trajectory is described by the
radius vector $X(X^i)$ and the velocity vector $\dot X ({\dot
X}^i)$; similarly, $x_{(r)}(x_{(r)}^i)$ and ${\dot x}_{(r)}({\dot
x}_{(r)}^i)$ are the radius vector and velocity vector of the
$r$th inerton. Let us modernise the Lagrangian (1) by introducing
the function $U_1$, which incorporates the intrinsic motion of the
particle and inerton ensemble
\begin{eqnarray}
    L=-M_0c^2 \Bigl\{ 1&-& \bigl[\dot X^2 + \sum_r U(X, \dot X;
     x_{(r)},  \dot x_{(r)})  \nonumber  \\
 &+&  \sum_r U_1(\Xi, \dot \Xi; \xi_{(r)},
 \dot \xi _{(r)})\bigr]/c^2 \Bigr\}^{1/2}.
 \label{2}
\end{eqnarray}

We suppose that the intrinsic motion of the particle is related
with the oscillation of the particle's centre-of-mass and is a
consequence of an asymmetric deformation of the particle volume at
the initial moment  at which the particle acquires a momentum.
Indeed, as the particle is considered to be solid and elastic
(like a drop), then the induction of the convace on one side must
automatically  result in the appearance of the convex on the other
side (Fig. 1). However the elasticity also implies the appearance
of the reverse force which restores the particle state. We will
examine the intrinsic motion of the particle in the same frame of
reference that the spatial motion. Because of this, one will
describe the deflected position of the particle centre-of-mass
along the particle trajectory  by the radius-vector $\Xi(\Xi^i)$
and depict the speed of its motion by the intrinsic-velocity
vector $\dot\Xi({\dot\Xi}^i)$; in a similar way, $\xi _{(r)}$ and
${\dot \xi}_{(r)}$ are the radius-vector and velocity of the
intrinsic motion of the {\it r}th inerton. It should be noted that
the point over the vectors $\dot X$ and $\dot \Xi$ means
differentiation with respect to the proper time {\it t} of the
particle which characterises the particle  along its trajectory;
by analogy, the point over the vectors ${\dot x}_{(r)}$ and ${\dot
\xi}_{(r)}$ implies differentiation on the proper time $t_{(r)}$
of the $r$th inerton.
\begin{figure}
\begin{center}
\includegraphics[scale=2.2]{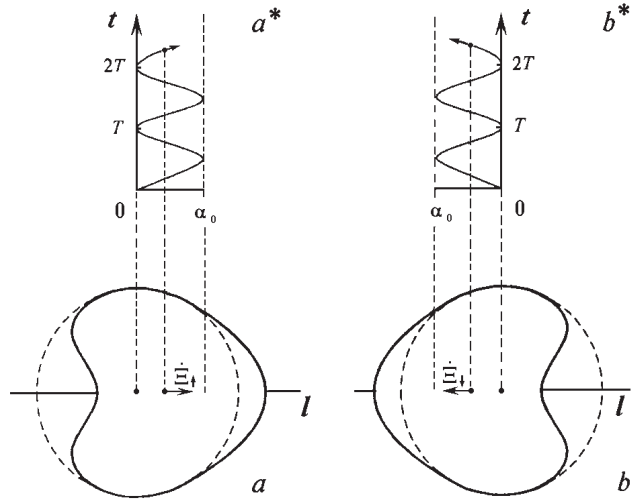}
\caption{The pulsation of the particle along its spatial
trajectory:  $a$ the convex is oriented in the direction of the
particle spatial motion and this orientation determines the
intrinsic-velocity vector $\dot \Xi_{\uparrow}$; $b$ the convex is
oriented against the particle spatial motion and this orientation
determines the intrinsic-velocity vector $\dot \Xi_{\downarrow}$.
Figures $a^*$ and $b^*$ schematically show the trajectory of the
particle centre-of-mass in a frame of reference connected with the
cell, which includes the oscillating particle ($\alpha_0$ is the
amplitude of these intrinsic oscillations).} \label{Figure 1}
\end{center}
\end{figure}

     In the intrinsic motion one of the two possibilities can be
realized: either the position of the particle's centre-of-mass is
displaced forward from the equilibrium one (Fig. 1$a$) or, on the
contrary, the position of the particle's centre-of-mass is
displaced backwards from the one (Fig. 1$b$). Then,  in  the
former case, the
 intrinsic velocity vector $\dot \Xi$ is aligned with the translational
 motion of the particle, i.e. parallel to the vector $\dot X$;  in
the latter case, it is opposing the  translational  motion, i.e.
 antiparallel to the vector $\dot X$.  The same is also true  for the
 vector of the intrinsic velocity of inertons.  In order to discern
 these two phase states of radial pulsations of the particle,
 let us use the index $\alpha$ for the intrinsic-velocity vector of
 the particle  and  of  the $r$th  inerton that defines  the
 intrinsic-velocity  projection  on  the  respective trajectories of
 the particle and the  $r$th inerton. One presumes that $\alpha =
 \uparrow$ for the former case (the  motion-oriented projection) and
$\alpha =\downarrow$ for the latter case (the motion-opposing
projection). Thus, hereinafter, we shall use the notations ${\dot
\Xi}_{\alpha}$ and ${\dot \xi}_{\alpha}$ for the intrinsic-velocity
vectors.

     Allowances can be made for the two possible phase states  of
 the particle in the Lagrangian if it  is  given  as  the  two-row
 matrix
\begin{equation}
{\cal L}=  {{\cal L}_{\uparrow} \choose {\cal L} _{\downarrow}}.
\label{3}
\end{equation}
Let us represent the function ${\cal L}_{\alpha}$ in the form
(compare with Ref. [32]):
\begin{equation}
{\cal L}_{\alpha}=-gc^2 \bigl\{ 1-[U^{\rm spat}+U^{\rm intr}_{\alpha}
]/gc^2 \bigr\}^{1/2},
\label{4}
\end{equation}
\begin{eqnarray}
&&U^{\rm spat}= g_{ij} \dot X^i(t) \dot X^j(t) + \sum_r \tilde
g_{ij}^{(r)} \dot x^i_{(r)}(t_{(r)}) \dot x^j_{(r)} (t_{(r)}) \ \
 \nonumber
\\
&&-\sum_r \frac {2\pi}{ T_{(r)}} \delta _{t-\Delta t_{(r)},
t_{(r)}} \Bigl[ X^i(t)\hat B_{ij} \dot x ^j_{(r)}(t_{(r)})+ \dot
X^i(t)\big\vert_{t=0} \hat B_{ij}x^j_{(r)}(t_{(r)}) \Bigr];
\label{5}
\end{eqnarray}
\begin{eqnarray}
&&U^{\rm intr} = g_{ij} {\dot \Xi}_{\alpha}^i(t) {\dot
\Xi}_{\alpha}^j(t) + \sum_r \tilde g_{ij}^{(r)}{\dot
\xi}^i_{\alpha(r)}(t_{(r)}){\dot \xi} ^j_{\alpha(r)} (t_{(r)})
  \nonumber        \\
&& \ \ \ \ \ \ \ \ \  -\sum_r \frac {2\pi}{ T_{(r)}} \delta
_{t-\Delta t_{(r)}, t_{(r)}} \Bigl[ \Xi^i_{\alpha}(t) \hat B_{ij}
\dot \xi^j _{\alpha(r)}(t_{(r)})+ {\dot
\Xi}^i_{\alpha}(t)\big\vert_{t=0} \hat
B_{ij}\xi^j_{\alpha(r)}(t_{(r)}) \Bigr];  \label{6}
\end{eqnarray}

            In (5) the three terms   describe
 the kinetic energies of the particle and of the inerton  ensemble
 and their interaction energy respectively. In (6) the three terms
 describe  the kinetic energies of  the
 intrinsic motion of the particle and of the inerton ensemble  and
 the energy of the intrinsic interaction between the particle  and
 the inerton ensemble accordingly. In (5) and  (6)
$g_{ij}$      are components  of
 the metric nonparametrised tensor created by the particle in  the
 three-dimensional space, along the particle trajectory \
$g_{ij}={\rm const} \ {\delta}_{ij}$; \   $g=g_{ij}\delta^{ij}$ is
the convolution of the tensor. The $r$th  inerton moves in the
field $g_{ij}$ and is characterized by the proper metric tensor
with the components ${\tilde g}^{(r)}_{ij}$ in the
three-dimensional space; these components describe a local
deformation of space in the neighborhood of the $r$th inerton (the
index $r$ is enclosed in parentheses in  order  to  distinguish it
from  the indices $i,\  j$, and $s$ which represent the vector and
tensor quantities). Along the trajectory of the $r$th inerton, the
tensor ${\tilde g}^{(r)}_{ij}$ is locally equal to \  const$\
{\delta}_{ij}$ \ approximately.\  $1/{T_{(r)}}$ is the frequency
of mutual collisions between the particle and the $r$th inerton,
hence, $T_{(r)}$ is the time interval elapsed from the moment of
emission of the $r$th  inerton  to  the moment of its absorption;
${\delta}_{t-{\Delta t_{(r)}}, t_{(r)}}$  is the Kroneker's symbol
that provides an agreement between the proper time $t$ of the
particle and  the proper time $t_{(r)}$  of the $r$th inerton and
$\Delta t_{(r)}$    is the time interval after the elapse of which
(starting from the  moment of the particle motion $t=0$) the
 particle emits the $r$th inerton.  Besides in (5) and (6) the
 notation
 \begin{equation} \hat B_{ij}=\sqrt{ g_{is}(\hat A^{-1}\tilde
    g_{sj}^{(r)})_0}, \label{7}
\end{equation}
is introduced where the operator ${\hat A}^{-1}$ takes inertons to
the trajectories distinct from  the  trajectory of the  particle
[32].

     Below we shall employ the approximation where the quantities
 ${\hat B}_{ij}$ are constant; for this purpose the proper time $t$ of
 the particle  should be regarded as the parameter proportional to
 the natural  $l$,\  \ $t \propto l$  \ ($l$   is the
trajectory length). In this case, the Euler-Lagrange equations for
the variables  ($X^i, {\dot X}^i$)  and  ($x_{(r)}, {\dot x}_{(r)}$)
 agree  with  the  respective  equations  of  extremals  for  the
 nonrelativistic  Lagrangian [32,33].
Similarly obtainable are the equations for the  intrinsic  variables \
(${\Xi}^i_{\alpha}, {\dot \Xi}^i_{\alpha}$)\  and \
(${\xi}^i_{\alpha ,(r)}, {\dot \xi }^i_{\alpha, (r)}$).
 \ As the proper time $t$  of the particle   is related
 with that of the $r$th inerton  by means of the relation
$t=t_{(r)}+{\Delta}t_{(r)}$, the equations of extremals  for  the
 intrinsic variables can be written through the parameter $t_{(r)}$.

    In the Euclidean space the Euler-Lagrange equations for  the intrinsic
variables of the Lagrangian (4)  take  the  form  (compare the
respective equations for space variables in Refs. [32, 33])
\begin{equation}
\ddot \Xi_{\alpha r}(t_r)+ \frac {\pi}{T_r} \frac {v_{0r}}{c}
\dot \xi_{\alpha r}^{\perp}(t_r)=0;
 \label{8}
\end{equation}
\begin{equation}
\ddot \xi^{\perp}_{\alpha r}(t_r)-\frac {\pi }{T_r} \frac
{c}{v_{0r}} \Bigl[\dot \Xi_{\alpha r}(t_r) - \dot \Xi_{\alpha
r}(t_r)\big\vert _{t_r=0 }\Bigr]=0; \label{9}
\end{equation}
\begin{equation}
\ddot \xi^{\parallel}_{\alpha r}(t_r)=0,\label{10}
\end{equation}
from this point on, the index $r$ is not parenthesis. In Eqs. (8) and
 (9) the intrinsic vectors ${\ddot \Xi}_{{\alpha}r}$
and ${\dot \Xi}_{{\alpha}r}$  are oriented parallel   or
 antiparallel to the $X \equiv X^1$ axis, along which the particle
 moves; ${\dot \xi}^{\perp}_{{\alpha r}}=\sqrt{({\dot \xi }^2_{{\alpha}
  r})^2 +({\dot \xi }^3_{{\alpha }r})^2}$    is   the
 intrinsic-velocity vector projection of the $r$th inerton that is
 oriented parallel or antiparallel to the axis which is normal to
 the  $X$ axis.  In this  case,  the projection  of the  intrinsic
 velocity of inertons onto the $X$ axis
is ${\dot \xi}^{\parallel}_{{\alpha}r}(t_r)=0$.

     For the sake of convenience, let us represent the intrinsic-velocity
vector for the particle and the $r$th inerton in the form
(hereinafter, we omit   the index $\perp$ at $\xi$)
\begin{equation}
\dot \Xi_{\alpha
r}(t_r)=e_{\alpha}\dot \Xi_r (t_r), \ \ \ \ \dot \xi_ {\alpha
r}(t_r)=e_{\alpha }\dot \xi_r (t_r),
\label{11}
\end{equation}
where the polarization symbol is inserted
\begin{equation}
e_{\alpha}= \cases {1  & if $\alpha=\uparrow $,  \cr
                    -1 & if $\alpha =\downarrow $. \cr}
\label{12}
\end{equation}
 As may be inferred from the form  of  Eqs.  (8)
 and  (9),  their solution essentially depends  on  the  quantity  of
 the  initial intrinsic velocity of the particle ${\dot \Xi}_{\alpha r}(t_r)
 {\vert}_{t_r=0}$.\  It seems reasonable to relate this value to
 the quantity of  the initial velocity of the particle's translational
 motion $v_o$ that is typical for the particle on the trajectory; we
 consider  the velocity of light $c$ as peculiar to
 inertons.  Then  the  initial velocity ${\dot \Xi}_r \big\vert_{t_r
 =0}$ is determined as
\begin{equation}
 \dot \Xi_r(t_r)\big\vert _{t_r=0}=v_{0r},
\label{13}
\end{equation}
 where $v_{0r}$ is the velocity of the particle at the moment
 of its collision with the $r$th  inerton [32]:
\begin{equation}
v_{0 r}=v_0
 [1-\sin(\pi r/2N)]; \label{14}
\end{equation}
 $N$ is the total number of the
 particle-emitted inertons (for $t_r \big\vert_{r=0}=t=0$ the quantity
 $v_{0r}\rightarrow v_0$).                                                         With regard for the above-mentioned, the solutions for  Eqs.
 (8) and (9) are easily obtainable:
$$
\dot\Xi_{\alpha r}=e_{\alpha}v_{0r}[1-\sin(\pi t_r/T_r)];
\eqno(15a)
$$
$$
\Xi_{\alpha r}=e_{\alpha}\Bigl[v_{0r}t_r + \frac {v_{0 r}T _r}{\pi}
[\cos(\pi t_r/T_r)-1]\Bigr];
\eqno(15b)
$$
$$
\dot \xi_{\alpha r} =e_{\alpha}c\cos(\pi t_r/T_r);
\eqno(16a)
$$
$$
\xi_{\alpha r}=e_{\alpha}{\Lambda_r\over \pi}\sin(\pi t_r/T_r).
\eqno(16b)
$$
 Solutions (15) and (16) for the intrinsic variables are true only
 within the time interval  $T_r= T(1-r/N)$ and they are identical in
 form to the solutions [33] for space variables
$X, \dot X$ \ and \  $x^{\perp}_r, {\dot x}^{\perp}_r$
(schematically the motion of the particle surrounded by its own
inertons is shown in Fig. 2). In (15) and (16) it is put $$
\lambda _r =v_{0r}T_r,\ \ \ \ \ \ \Lambda_r=cT_r. \eqno(17) $$
\begin{figure}
\begin{center}
\includegraphics[scale=2]{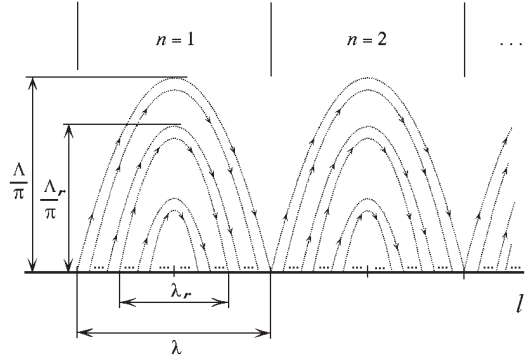}
\caption{The schematic representation of the motion of the
particle surrounded by inertons along the particle trajectory
$l$.} \label{Figure 2}
\end{center}
\end{figure}

     Having regard to
$$ \dot X\vert _{t=nT}=v_0 \eqno(18) $$
where $n  =  1, \ 2, \ 3,\
...$ \  is  the  number  of the time period $T$ of collisions  in
accordance with which the quantity of the particle velocity
resumes its initial value,  one can present [33] the solutions for
$X$ and $\dot X$ as a function of the proper time $t$ of the
particle :
  $$ \dot X(t)=v_0(1-\vert \sin(\pi t/T)\vert);
\eqno(19a) $$ $$ X(t)=v_0t + \frac {v_0 T}{\pi} \Bigl\{
(-1)^{[t/T]} \cos(\pi t/T) -\bigl(1+2[t/T]\bigr) \Bigr\};
\eqno(19b)
 $$
here the notation $[t/T]$ means an integral part of the integer
$t/T$. By analogy we can write the solutions for the intrinsic
variables of the particle  that  are also  true  along its  entire
spatial trajectory (Fig. 3):
   $$ \dot
\Xi_{\alpha}(t)=e_{\alpha}v_0 (1-\vert\sin(\pi t/T)\vert);
\eqno(20a)
  $$
  $$ \Xi_{\alpha}(t)=e_{\alpha} \Bigl\{ v_0 t +
\frac {v_0 T}{\pi} \bigl\{ (-1)^{[t/T]} \cos(\pi
t/T)-\bigl(1+2[t/T] \bigr) \bigr\} \Bigr\} \eqno(20b)
$$
\begin{figure}
\begin{center}
\includegraphics[scale=2]{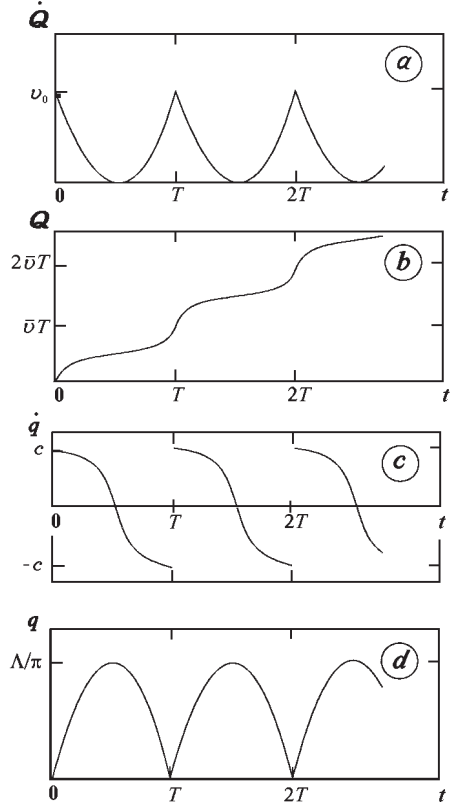}
\caption{The schematic representation of the solutions of Eqs.
(19) and (20) for the particle ($a, \ b$) and Eqs. (23) for the
inerton cloud ($c,\  d$).  Here $Q \equiv \{ X; \ \Xi_{\alpha}
\}$, $\dot Q \equiv \{ \dot X; \ {\dot \Xi}_{\alpha} \}$ and $q
\equiv \{ x; \ \xi_{\alpha} \}$, $\dot q \equiv \{ \dot x; \ {\dot
\xi}_{\alpha} \}$. Note that the value $\bar v T$ shown in the
figure {\it b} can be interpreted as a mean de Broglie wavelength
$<\lambda>$ of the particle along its trajectory. } \label{Figure
3}
\end{center}
\end{figure}

\section{\bf Hamiltonian of a "bare" particle}

\vspace{2mm} \hspace*{\parindent}
      Further analysis is performed  within  the  framework  of  the
 Euclidean space viewing the inerton ensemble as a  single  object,
 i.e. the inerton cloud. Let  masses of the  particle  and  the
 cloud at rest be equal to $M_0$ and $m_0$ respectively. Assume that
 the  particle  trajectory  runs  along  the $X$ axis  and let
 $x$ is the distance between the cloud and the particle, then $\dot
 x$ \ is the velocity of the cloud in a frame of reference connected
with the particle.  The intrinsic coordinate
$\Xi_{\alpha}=e_{\alpha}\Xi$  defines the location  of the
 centre-of-mass of the pulsating particle along  the particle's
 trajectory; \ the intrinsic-velocity vector
 ${\dot {\Xi}}_{\alpha}=e_{\alpha}{\dot {\Xi}}$ of the particle
sticks out of the point of the particle's  centre-of-mass location.
Similarly for the inerton cloud:  $\xi_{\alpha}=e_{\alpha} \xi$,
  where $\xi$   is the   displacement of the
cloud's pulsating centre-of-mass from the  equilibrium position  and
   $\dot\xi_{\alpha}=e_{\alpha}\dot\xi$, where $\dot\xi$
  is the  intrinsic-velocity vector that sticks out of
the position of the cloud's centre-of-mass, i.e. $\dot \xi$ is the
velocity vector of the pulsation.  Both the particle's ($X,
 {\dot X}; \Xi, \dot\Xi$) and cloud's ($x,\dot x; \xi,
 \dot\xi$)  parameters are functions of the proper time $t$
 of the particle. Thus  instead of  (3) -- (6), we proceed from the
Lagrangian $L=\Vert L_{\alpha} \Vert$  where
$$
   L_{\alpha}=-M_0c^2 \Bigl\{1  \ \ \ \ \ \ \ \ \ \ \ \ \ \ \ \ \ \ \ \
\ \  \ \ \ \ \ \ \ \ \ \ \ \ \ \ \ \ \ \ \ \ \ \ \ \ \ \ \ \ \ \ \
\ \ \ \ \ \ \ \ \ \ \ \ \ \ \ \  \ $$ $$ - \frac 1{M_0c^2}
\Bigl[M_0\dot X^2 + m_0\dot x^2-\frac {2\pi}{T}\sqrt{m_0M_0}(X\dot
x + v_0x)\Bigr] \ \ \ \ \ \ \ \ \ \ $$ $$ -\frac 1{M_0c^2}
\Bigl[M_0\dot \Xi_{\alpha}^2+m_0{\dot \xi}_{\alpha}^2- \frac
{2\pi}{T}\sqrt{m_0M_0}(\Xi_{\alpha}\dot \xi_{\alpha}+
 e_{\alpha}v_0\xi_{\alpha})\Bigr] \Bigr\}^{1/2}. \eqno(21)
 $$
 Here $1/T$ is the frequency of collisions between
 the particle and the inerton cloud, $v_0$ is the initial particle
 velocity.

    If the parameter of running for the functional of extremals
$$ S_{\alpha}=\int_A^B L_{\alpha} d t \eqno(22) $$ is thought of
as being natural, \  $t \propto l$  ($l$  is the $AB$   curve
length or in other words the particle path), solution to the
Euler-Lagrange  equations for both spatial and intrinsic variables
of the Lagrangian (21) are readily obtainable. The
particle-describing solutions for the variables \  ${\dot X}, X$ \
and \  ${\dot \Xi}_{\alpha}, \Xi_{\alpha}$ \ agree with the
respective expressions (19) and (20).  For solutions  for the
spatial variables $\dot x$ and $x$ of the inerton cloud see  Ref.
[33] and the solutions for the intrinsic variables of the cloud
coincide in form with (16) where the sine and cosine should be put
into the modules sign  and   the index $r$ should be dropped. In
other words, the solutions for the cloud have the form
 $$ q=\frac
{\Lambda}{\pi} \vert  \sin (\pi t/T) \vert; \eqno(23a)
 $$
  $$ \dot
q = (-1)^{[t/T]} c \cos(\pi t/T) \eqno(23b)
  $$
 where $q\equiv\{x;\ \xi_{\alpha} \}$ and $\dot q \equiv \{ \dot x; \ {\dot
\xi}_{\alpha} \}$. Note that one can choose the proper time of the
particle along its trajectory in the form of $t=l/ \bar v$ where
$\bar v$ is the mean velocity  of the particle along its
trajectory. Owing to expression (19$a$) one finds
 $$ \bar v  =
\frac 1{T} \int\limits^T_0 \dot X d t  = v_0 (1- 2/ \pi).
\eqno(24)
 $$
With regard for expression (24) the solution of equations of
motion for the particle ((19) and (20))  take the symmetric form.

    The substitutions
$$
{\dot {\tilde x}}={\dot x}-\frac {\pi}{T}\sqrt{M_0/m_0} \ X,
\eqno(25)
$$
$$ {\dot{\tilde \xi}}_{\alpha}={\dot \xi}_{\alpha} - \frac
{\pi}{T}\sqrt{M_0/m_0} \  \Xi_{\alpha}
\eqno(26)
$$
reduce the Lagrangian (21) to the form
$$
L_{\alpha}=-M_0c^2
\Bigl\{ 1 \ \ \ \ \ \ \ \ \ \ \ \ \ \ \ \ \ \ \ \ \ \ \ \ \ \ \ \ \ \ \
\ \ \ \ \ \ \ \ \ \ \ \  \ \ \ \ \ \ \ \ \ \ \ \ \ \ \ \ \ \ \ \ \ \ \
\ \ \ \ \ \ \ \ \ \ \ \  $$
$$
-\frac
1{M_0c^2}\Bigl[M_0\dot X^2-M_0\omega^2 X^2+ m_0\dot {\tilde x}^{\
2}-2\omega \sqrt{m_0M_0} \  v_0 x\Bigr]   \ \ \ \ \ \ \ \ \ \
$$
$$-\frac 1{M_0c^2}\Bigl[M_0 \dot\Xi^2_{\alpha}-M_0\omega^2
\Xi^2_{\alpha}+m_0\dot{\tilde\xi}^{\ 2}- 2\omega \sqrt{m_0M_0}
\ v_0e_{\alpha}\xi_{\alpha}\Bigr] \Bigr\}^{1/2}
\eqno(27)
$$
where the notation
$$
\omega ={2 \pi}/ 2T
\eqno(28)
$$
is inserted.

    Let us build  up the Hamiltonian function appropriate to the
Lagrangian (27):
$$
H_{\alpha}={\dot X} \partial L_{\alpha}/\partial \dot X +
\dot{\tilde x} \partial L_{\alpha} / \partial \dot{\tilde x} +
\dot\Xi_{\alpha} \partial L_{\alpha}/\partial \dot\Xi_{\alpha}+
\dot{\tilde\xi}_{\alpha}\partial L_{\alpha}/\partial \dot{\tilde\xi}
_{\alpha} - L_{\alpha}.
\eqno(29)
$$

Substituting $ L_{\alpha}$ from (27) into (28), we derive
$$
H_{\alpha}= M\omega^2X^2 + M\omega^2\Xi^2_{\alpha}+
2\omega \sqrt{mM_0} \ v_0
(x + e_{\alpha}\xi_{\alpha}) + Mc^2
\eqno(30)
$$
where the designations
$$
M=M_0/\sqrt{1-v_0^2/c^2}, \ \ \ \ \ \ m=m_0/\sqrt{1-v^2_0/c^2}
\eqno(31)
$$
are entered (the derivation of (30) presupposes that in the Lagrangians
(21) and (27), the radical \ $ \sqrt{...}$ \ is constant along
the particle trajectory and equal to $\sqrt{1 - v_0 ^2 / c^2}$).

     On the other hand, the momentum is determined as equal  to
${\partial L}/{\partial {\dot Q}}$ where $\dot Q$ is
the canonical velocity. The Lagrangian (27) is a function of the four
velocities, so we can introduce their four respective momenta:
$$
p=\partial L_{\alpha}/\partial \dot X =M\dot X;
\eqno(32a)
$$
$$
\tilde p \  = \partial L_{\alpha}/\partial \dot{\tilde x} \ \ = m\dot
{\tilde x}; \eqno(32b)
$$
$$
\pi_{\alpha}=\partial L_{\alpha}/\partial \dot\Xi_{\alpha}
=M\dot\Xi_{\alpha};   \eqno(32c)
$$
$$
\tilde \pi_{\alpha} =\  \partial L_{\alpha}/\partial \dot{\tilde\xi}
_{\alpha}=m\dot {\tilde\xi}_{\alpha}.
\eqno(32d)
$$
Expressions (32) permit the Hamiltonian (29) to be represented as
$$
H_{\alpha}=p^2/M+{\tilde p}^{\ 2} /m + \pi^2_{\alpha}/M +
{\tilde\pi}^{\ 2}_{\alpha}/m +(M_0c)^2/M. \eqno(33)
$$

    Both the Hamiltonians, (30) and (33), are identical and their
superposition makes it possible  to write the Hamiltonian  for the
particle -- inerton  cloud  system as $$
 H_{\alpha}=H^{\rm part}_{\alpha}+H^{\rm cloud}_{\alpha}
      +H^{\rm renorm}_0
\eqno(34)
$$
where the effective Hamiltonians of the particle  and
the cloud are respectively equal to
$$
        H^{\rm part}_{\alpha}=p^2/2M
         +M\omega^2X^2/2 +\pi_{\alpha}^2/2M
         +M\omega^2 \Xi^2_{\alpha}/2;
\eqno(35)
$$
$$
  H_{\alpha}^{\rm cloud}={\tilde p}^{\  2}/2m
  +{\tilde\pi}^{\ 2}_{\alpha}/2m +2\omega \sqrt{mM_0} \ v_0
(x + e_{\alpha}\xi _{\alpha}). \eqno(36) $$ The renormalized
energy of the particle  at rest  involved in (34) is $$
 H^{\rm  renorm}_0 = (M_0c)^2/2M +Mc^2/2.
\eqno(37)
$$

     The Hamiltonian (35) governs  the behaviour  of a "bare"
relativistic particle in the phase space, the first two terms
describe spatial oscillations of the particle relative to the
inertia centre of the particle and inerton cloud and the last two
terms represent particle intrinsic oscillations. Following papers
[32,33] we now pass from the function (35) to the Hamilton-Jacobi
equations for the shortened action  (the spatial $S_1$ and
intrinsic $\Sigma _{\alpha 1}$ ) of the particle:
     $$
(\partial S_1 /\partial X)^2/2M + M \omega^2X^2/2 =E, \eqno(38)
     $$
    $$ (\partial \Sigma_{\alpha 1}/ \partial \Xi_{\alpha})^2/2M +
M\omega^2 \Xi^2_{\alpha}/2 =E_{\alpha}. \eqno(39)
    $$
    Here the constants $E$ and $E_{\alpha}$ are the respective
energies of the spatial and intrinsic motions and these quantities
should be identified with the initial energy of the oscillator,
i.e. with the kinetic energy equals  $M v_0^2 /2$. So $$
E=E_{\alpha} = Mv_0^2 / 2. \eqno(40) $$ From Eqs. (38) and (39)
one deduces (see, e.g. ter Haar [44], Goldstein [45]):
   $$
X=\sqrt{2E/M\omega^2} \sin{\omega t}, \eqno(41) $$
   $$
\Xi_{\alpha}=\sqrt{2E_{\alpha}/M\omega^2} \sin{\omega t}.
\eqno(42)
   $$

       Periodicity in the motion of the particle  allows for passing
on to the action -- angle variables in Eqs. (38) and (39) that
describe its motion. For the increment of the action within the
oscillation period $2T$ we derive
  $$ J=\oint pdX =E\times 2T;
\eqno(43)
  $$
  $$ \iota_{\alpha}=\oint\pi_{\alpha} d
\Xi_{\alpha}=E_{\alpha}\times 2T. \eqno(44)
  $$
  With regard to (40) and (17), we find from (43) and (44):
  $$
J=Mv_0 \times v_0 T =p_0 \lambda; \eqno(45)
  $$
  $$
\iota_{\alpha}=Me_{\alpha}v_0\times e_{\alpha}v_0T=\pi_{\alpha
0}\lambda _{\alpha}.\eqno(46)
 $$

According to (40), the quantity $\iota _{\alpha}$ determined in
(44) does not depend on $\alpha $; hence, assuming that $J = \iota
= h$ where $h$ is Planck's constant, we obtain from (45) and (46)
the de Broglie relation
 $$ h /\lambda =Mv_0 \eqno(47) $$
where, however, $\lambda$ is not a certain wave but is a spatial
period or, equivalently, the oscillation amplitude of the moving
particle. As $2T$ is the cyclic oscillation period, by introducing
the frequency $\nu = 1/2T $ into (43) and (44) in view of the
equality $J = \iota = h$  we obtain another basic relation in
quantum mechanics: $$
    E=h \nu;   \eqno(48)
$$
the constant $E$ being adequate  to the initial kinetic energy
of the particle.

\vspace{4mm}

\section{\bf Wave equation for spin}

\vspace{2mm} \hspace*{\parindent}
      As is known, de Broglie [46], the relations (46) and (47)
admit the wave equation for space variable $\vec X$ to be written as
$$
 \Bigl( \frac {\hbar^2}{2M}\nabla^2_{\vec X} - E \Bigr) \psi =0.
\eqno(49)
$$
But the same relations (47) and (48) hold for the intrinsic variable
${\vec {\Xi}}_{\alpha}$ as well, so in this case, too, we could formally
insert the wave equation.

       However, in Eq. (49) the  eigenvalue $E$ describes
the particle spectrum.  This quantity is observable, i.e., it is
immediately measured by the instrument. The quantity $| \psi |^2$
governs some mean coordinate and momentum of the particle. For a
detailed up-to-date interpretation of the $\psi$-function one
refers to the  review of Sonego [47]. Besides Oudet [25] has shown
that components $\psi_s$ ($s=\overline {1,\ 4}$) in Dirac's
equation might be connected with the exchange by "grains" between
the electron and its field (the "grain" is an element of the total
electron mass). Evidently, in our approach the role of those
grains is played by inertons.  The present theory gives an
additional information.  It points to a spatial vicinity of the
$\psi-$function extension around the particle. It is obvious that
the vicinity is defined  by the dimensions of the inerton cloud:
$\lambda$ along the particle path and $2\Lambda/\pi$ in
transversal directions.

     At the same time, when measuring the
 quantum system undisturbed under a certain external influence,
its intrinsic variables are not manifested. Therefore, in this
case, the eigenfunction ($\chi _{{\alpha} 0} $)  and eigenvalue
($\varepsilon _{{\alpha} 0}$) that describe intrinsic degrees of freedom
should be identically equal to zero, the intrinsic operators of
the coordinate ${\hat{\vec {\xi} }} _{\alpha}$ and of the momentum
${\hat{\vec {\pi} }}_{\alpha} = - i \hbar {\nabla}_{{\vec {\xi}}_{\alpha}}$
being potential. But  the external field being superimposed on the system,
 the intrinsic variables can fall into engagement with the field and,
as a consequence, will be able to become explicit, when measuring.

     In this way, the spatial momentum operator in the magnetic field
with the vector  potential $\vec A$ is $({\hat{\vec p}}-{e \vec A})$.
 Analogously, the intrinsic momentum operator ${\hat {\vec \pi}}_{\alpha}$
 should apparently be substituted for the generalized one
$\hat{\vec \Pi} _{\alpha}=(\hat{\vec \pi} _ {\alpha} - {e \vec A })$;
recall that orientations of  basis vectors $e^1, \ e^2$, and $e^3$ for
spatial ($X^1, \ X^2, \ X^3$)  and intrinsic
($\xi ^1 _{\alpha}, \xi ^2 _{\alpha}, \xi ^3 _{\alpha}$)
coordinates of the particle coincide. If the induction
 of magnetic field has only one component  ${\cal B}$  aligned with the axis
 $Oe^3$, Eq. (49) in the central-symmetric potential $V(\vec X)$ is known
 (see, e.g. Sokolov {\it et al.} [48]) to reduce to the form
$$
\Bigl(E-V(\vec X)-{\hat{\vec p}}^{\ 2}/2M + e{\cal B}\hat L_3/2Mc\Bigr)
\psi =0
\eqno(50)
$$
where $\hat L_3$ is the operator component of the moment of momentum
along the $Oe^3$ axis ($L_3 = \hbar {\bar m}$,  where  $\bar m = 0,
\pm 1, \pm 2, \ ...$).

     An apparent manifestation of the intrinsic motion of the free particle
in the field of magnetic induction $\cal B$ implies
the guidance of the nonzero intrinsic
 eigenvalue $\varepsilon _{\alpha}$ and eigenfunction $\chi_{\alpha}$.
As a result, in this case, we have the right to write  the equation
$$
  \Bigl({\hat{\vec\Pi}}^{\ 2}_{\alpha}/2M - e_{\alpha}\varepsilon_{\alpha}
    \Bigr) \chi_{\alpha}=0.
\eqno(51)
$$
The eigenvalue shows to what extent the particle energy is changed
when the intrinsic degree of freedom interacts with the
induction $\cal B$, that is why the quantity $\varepsilon _{\alpha}$
being either positive or negative. However, in view of a specific
construction of Eq. (49) (according to Eq. (49), in Eq. (50) the
eigenvalue $E>0$), we similarly introduce only a positively
determined parameter $e_{\alpha}{\varepsilon _{\alpha}}$ into Eq. (51)
where $e_{\alpha}$ is defined in (12). Allowing that the induction
$\cal B$ is aligned solely with the $Oe^3$ axis, instead of Eq.
(51) we obtain
$$ \bigl({\hat\Pi_{\alpha 1}}^2/2M + {\hat\Pi_{\alpha
2}}^2/2M - e_{\alpha}\varepsilon_{\alpha} \bigr) \chi_{\alpha}=0.
\eqno(52)
$$
As
$$
 [\hat\Pi_{\alpha 1},\hat\Pi_{\alpha 2}]_{\_}=ie\hbar{\cal B},
\eqno(53) $$ dimensionless variables  ${\hat Q}_{\alpha}$ and
${\hat P}_{\alpha}$ determined according to the rule $$
\hat\Pi_{\alpha 1}=C\hat Q_{\alpha},\ \ \ \  \hat\Pi_{\alpha 2}=
C\hat P_{\alpha};\ \ \ \ C=(e\hbar {\cal B})^{1/2} \eqno(54) $$
can be entered. It thus follows from commutation (53) that ${\hat
Q}_{\alpha}$ is the generalized dimensionless coordinate operator
and ${\hat P}_{\alpha}= i{\partial}/ \partial {\hat Q}_{\alpha}$
is generalized dimensionless momentum operator. With the new
variables Eq. (52) is rewritten as follows: $$ \frac {e\hbar{\cal
B}}{2M} \Bigl( \hat P^2_{\alpha}+\hat Q^2_{\alpha} \Bigr)
\chi_{\alpha}=e_{\alpha}\varepsilon_{\alpha}\chi_{\alpha}.
\eqno(55) $$

       It is obvious that Eq. (55) is the harmonic oscillator equation,
so we can find both its eigenfunction and eigenvalue. If chosen
out of a variety of solutions is only the one characterized by the
smallest magnitude of $e_{\alpha}\varepsilon _{\alpha}$, the
solution to Eq. (55) for this lowest level takes the form $$
\chi_{\alpha}=\pi^{-1/4}\exp[-\hat Q^2_{\alpha}/2]\equiv
\pi^{-1/4}\exp \Bigl[-\frac {(\hat{\pi}_{\alpha 1}-eA_1)^2}
{2e\hbar{\cal B}} \Bigr]; \eqno(56a) $$ $$
\varepsilon_{\alpha}=e_{\alpha}e\hbar{\cal B}/2M \eqno(56b) $$
(regard is made here that $e^2_{\alpha}=1$). As is seen from (56),
with $\cal B \rightarrow $0, the eigenfunction $\chi _{\alpha}$
and the eigenvalue $\varepsilon _{\alpha}$  tend to zero, as they
must. With allowance for $e_{\alpha} = \pm 1$, automatically
obtainable from $(56)$ is the expression for  $\varepsilon
_{\uparrow (\downarrow)}$ in the representation of the so-called
operator of the spin projection onto the axis $Oe^3$:
  $$
\varepsilon_{\uparrow (\downarrow)}= \frac {e \cal B} {M} {\hat
S}_3 \eqno(57)
  $$
where, as is known, the eigenvalues of this operator
 $$
S_{\uparrow (\downarrow) 3}=\pm\hbar/2. \eqno(58)
 $$

        Correction (57) to the particle energy which is due to the
intrinsic degree of freedom should be inserted into the total
particle spectrum, i.e. in Eq. (50) the eigenvalue should be
renormalised, $E \rightarrow E + e {\cal B} S_ {\uparrow
(\downarrow) 3}/M$.\  The  renormalised equation goes into the
Pauli equation. They differ only in the mass quantity: the
relativistic mass of the particle $M$ enters Eq. (50) but only the
mass at rest $M_0$  appears in the Pauli (and Schr\"odinger)
equation. However, if the terms proportional to $v_0 /c$ are
accounted for, Eq. (50) and the Pauli equation coincide.

\vspace{4mm}

\section{\bf Discussion}

\vspace{2mm} \hspace*{\parindent}
    Based on the total Hamiltonian written, for instance, as (33)
 the total Hamiltonian of the particle in the neglect
 of the availability of inertons appears as
 $$
H_{\uparrow (\downarrow)}^{\rm part. tot}=\vec p^{\ 2}/M +
\vec\pi_{\uparrow (\downarrow)}^{\ 2}/M + (M_0c)^2/M.
\eqno(59)
 $$
Since the total energy of the particle
$E_{\rm tot}=E_{{\uparrow}(\downarrow) {\rm tot}}=Mc^2$, (59) can
be represented in the form of
 $$
H_{\uparrow (\downarrow)}^{\rm part. tot}= \sqrt{
        c^2\vec p^{\ 2} +c^2\vec\pi^{\ 2}_{\uparrow (\downarrow)}
          + M_0^2c^4}.     \eqno(60)
 $$
If in (60) we pass on to the operators and linearise the Hamiltonian
(60) over $\hat {\vec p}$, the intrinsic momentum operators
${\hat {\vec {\pi}}}_{\uparrow (\downarrow)}$ have been excluded,
we derive the Dirac Hamiltonian
 $$
\hat H_{\rm Dirac}=c\hat{\vec{\alpha}} \hat{\vec p}
+{\hat{\varrho}}_3M_0c^2.
\eqno(61)
 $$
At this point, information on the operators ${\hat {\vec {\pi}}}_
{\uparrow (\downarrow)}$ immediately goes into the $\hat {\vec {\alpha}}$-
matrix.

      In  the relativistic quantum theory, ascribed to the particle is the
frequency which equals, according to de Broglie,
  $$ \nu_{\rm
rel}=\sqrt{p^2c^2 + M^2_0c^4} /h \eqno(62)
 $$
and it is precisely this frequency that characterises the spectrum
of the wave-particle in the Dirac wave equation
   $$ (i\hbar\partial
/\partial t - \hat H_{\rm Dirac} ) \psi_{\rm Dirac}=0. \eqno(63)
   $$
At the  same time the wave equations obtained in the previous
section refer to the particle as a corpuscle which is defined only
by the kinetic constituent $Mv_0^2/2 $ of the total energy of the
particle $Mc^2$ and these equations result from the established
ratio between the frequency $\nu$, amplitude $\lambda$ and initial
velocity $v_0$ of the oscillating particle: $\nu =
v_0/2{\lambda}$. In as much as the energy of the particle at rest
$M_0c^2$ is a peculiar intrinsic potential energy, it is
reasonable for our model that this energy does not displace itself
in the particle spectrum. So in what manner can the hidden energy
$M_0c^2$ make itself evident in an explicit form? We shall make an
effort to answer this question, consideration being given to the
particle-surrounding vacuum medium.

       Phase transitions in the vacuum are an urgent problem of
the field theory. In the vacuum the phase transition presumably
occurs with the development of singularity (see, e.g. Ginzburg
[49]) to which the equation of the state $$ {\cal E}_{\rm vac}=
-{\cal P}_{\rm vac} \eqno(64) $$ is employed (${\cal E}_{\rm vac}$
and ${\cal P}_{\rm vac} < 0$ are  respectively the energy density
and vacuum pressure in the region of singularity). In the present
model the particle is practically a point defect in space
structure. So it is logical that elastic space gives a linear
response to such a defect  when space suffers an elastic
deformation within some radius $R_{\rm sing}$. This deformed-space
region should evidently be regarded precisely as a space
singularity. It is reasonable to suggest in view of the response
linearity that the singularity region radius far exceeds the size
of the particle $R_{\rm part}$ which is presumably $\leq
 10^{-28}$cm. The response linearity provides an equality between
the energy of the particle at rest $M_0c^2$ and  the space
singularity region energy $E_{\rm sing}$, i.e.
 $$
 E_{\rm sing} - M_0c^2 = 0. \eqno(65)
 $$
  Thus the singularity region, i.e., the deformation
coat which developed around the particle [32-34], is similar to a
shell that shields the particle from degenerate space.

       In the author's approach  space is thought to be a
cellular structure where  each cell is occupied by a
superparticle; besides, it has been noted that the deformation of
a relatively wide region of space involves an induction of mass in
superparticles in this region (recall that we relate the emergence
of mass in the particle to the change in size in the degenerate
superparticle). Therefore, the particle-containing space
singularity region has a deformed cellular structure and in each
cell the superparticle possessed mass. Hence it may be inferred
that the given region should be seen as an ordered crystalline
structure. But a crystal is described by the vibrating energy of
its sites and in the present case of the space crystalline
singularity region  the role of sites is played by massive
superparticles. In the crystallite singularity vibrations of all
superparticles are cooperated and their total energy is quantized
(see Appendix). So Eq. (65) is reduced to the form
  $$
\hbar\omega_{\vec k_0} =M_0c^2 \eqno(66)
  $$
where $|{\vec k}_0|\equiv k_0 = {2\pi} /{\tilde {\lambda}}_0$  is
the wave number and ${\omega}_{{\vec k}_0} =ck_0$ is the cyclic
frequency of oscillator in the $\vec k$-space (the quantity
${\tilde {\lambda}}_0$ is the amplitude of this oscillator and is
given by crystallite singularity size). It is of interest that
according to (66)
  $$
\tilde \lambda_0 =M_0c/h \eqno(67)
  $$
and, as is obvious, ${\tilde {\lambda}}_0$  coincides with the
Compton wavelength of this particle. But the Compton wavelength
characterises an effective size of the particle at its scattering
by the photons, and, as may be seen, quite a reasonable
explanation is invoked in our model to account for this parameter.
Moreover, as we identify the deformation potential of the particle
with its gravitational potential, the quantity ${\tilde
{\lambda}}_0 /2$ should be treated as the limiting action radius
of the particle's gravitational field or, in other words, the
gravitational radius of the particle at rest. By the way, the
quantity ${\tilde {\lambda}}_0$ plays also a crucial role in the
orthodox relativistic quantum theory because  the value ${\tilde
{\lambda}}_0$ determines a minimum size for which the concept of
Dirac field still applicable.

       The moving particle has the total energy equals to
$Mc^2$, that is why a change in energy of the singularity region
will follow the particle energy alteration:
  $$ \hbar \omega_{\vec
k_{v_0}}=Mc^2 \eqno(68)
  $$
where the quantity $k_{v_0}$ is defined by the amplitude ${\tilde
{\lambda}}_{v_0}$ of another oscillator from the $\vec k$-space. A
comparison between (66) and (68) shows that as the particle
develops the velocity $v_0$, the singularity region along the
vector $\vec {v_0}$ decreases in size: ${\tilde {\lambda
}}_{v_0}={\tilde {\lambda}}_0 \sqrt{1-v_0^2/c^2}$.  The
singularity region travels together with the particle. The region
migration occurs due to the hopping the mass from the region's
massive superparticles to the nearest degenerate ones by stages
(the velocity of adjustment of superparticles $c>v_0$, i.e. the
singularity region always keeps pace with the particle).

    We now focus once again upon the nature of the motion
of the particle-emitted inertons. The inerton as an elementary
excitation of space is determined [32-35] as the deformation (the
size variation) of a superparticle, the initial size of
surrounding superparticles being retained. This definition is
evidently true to both the degenerated space and the crystallite
singularity. The inerton migrates from the  $r$th to $(r+1)$th
superparticle by a relay mechanism just as Frenkel excitons in
molecular crystals. The inerton motion is described by the two
components, the former being longitudinal and the latter
transverse relative to the particle trajectory. Lengthwise, the
inerton velocity is of the order of the quantity $v_{0r}$ (14) and
owes essentially to the momentum transmitted by the particle.
Crosswise, the  motion of inertons is in principle of different
nature: their migration is similar to the movement of the standing
wave profile in the string. Indeed, when moving, the particle
deforms, or excites a superparticle it encounters, in the
direction perpendicular to the particle's trajectory (by virtue of
the operator (7)). But superparticles elastic, so each $r$th
superparticle regains its original state, i.e. its size  and yet,
in doing so, its excitation is initially transmitted by a relay
mechanism deep into the space singularity region and thereafter
into the degenerate space. The motion of such quasi-particle is
described by the equation
  $$
m_rd^2x_r/dt^2 = -\gamma x_r. \eqno(69)
  $$
At the point maximum removed from the particle, that is
${x_r}{\vert}_{\rm {\max}}={\Lambda}_r$,  the kinetic energy $m_r
{{\dot x}_r}^2/2$ of the $r$th inerton passes into the potential
energy. At this point the inerton has no  mass: the superparticle
embedded at this point has no deformation (its size is the same as
nearest superparticles). However, here appears a local deformation
of space: the corresponding superparticle is maximally displaced
from its equilibrium positions in degenerate space in the
direction from the particle.  Consequently, this superparticle
experiences an elastic response (in the direction to the particle)
from the side of the whole space, which leads to the inerton
migration back to the particle. Inertons penetrate the boundary
between the crystallite and the degenerate space unobstructed,
without scattering as their motion is practically normal to the
surface of this boundary.

        The inerton cloud oscillation  amplitude \ $\Lambda/\pi$ \ is
connected  with the de Broglie wavelength (the spatial oscillation
amplitude)\ $\lambda$ \ of the particle by  the relationship [32]
(compare also two expressions in (17))
  $$
\Lambda = \lambda c/v_0. \eqno(70)
  $$
A similar connection exists between $\lambda$ and ${\tilde
{\lambda}} _{v_0}$: while comparing the formulas $\lambda =h/Mv_0$
for the particle and \ ${\tilde {\lambda}}_{v_0} =h/Mc$ \ for the
singularity region, we derive
  $$
\lambda =\tilde\lambda_{v_0}c/v_0.  \eqno(71)
  $$
Then from (70) and (71) one deduces the very interesting
relationship:
 $$
\Lambda = \tilde\lambda_{v_0}c^2/v_0^2.  \eqno(72)
 $$

    It follows from relation (72) that with $v^2_0/c^2 \ll 1$,
the inerton cloud amplitude is much superior to the size of the
singularity region, i.e.  $\Lambda \gg {\tilde {\lambda}}_{v_0}$.
In this case the inerton cloud  that surrounds the particle
governs its motion, as already at a distance of $\sim \Lambda $
from the particle the cloud suffers obstacles and conveys
pertinent information to the particle and this is the easiest
explanation of the particle diffraction phenomenon. It may be seen
that such motion is close to the L. de Broglie "motion by
guidance" [50] which he related to a constant intervention  of a
subquantum medium. Hence within $v^2_0 /c^2 \ll 1 $,  while
measuring the coordinate and/or the momentum of the particle along
the direction of its movement the instrument records the inerton
cloud that transfers the same kinetic energy along the particle
trajectory as the particle does, $Mv_0^2/2$. Thus the singularity
region, when measured, is apparently not displayed in a so called
nonrelativistic approximation, therefore in the present case it is
appropriate to use the Schr\"odinger and Pauli equations in order
to analyse the particle behaviour.

   In the so called relativistic limit, $v_0 \rightarrow c$,
it is evident from (72) that $\Lambda \approx {\tilde {\lambda }}_{v_0}$
and this  implies that  the inerton cloud  is  virtually  completely
closed in the singularity region (or, in other words, in the deformation
coat). Because of this, when measuring the coordinate  or the energy of
the particle along the direction of its motion, the instrument
will register the entire moving singularity region. But the total
energy $\hbar {\omega}_{{\vec k}_{v_0}} $  of the latter exceeds the
kinetic energy $Mv_0^2 /2$ of the inerton cloud and, consequently,
in this instance, the energy of the particle at rest $M_0 c^2$ will
explicitly reveal itself as well.

      As indicated above, the Dirac theory formally ascribes
the frequency $\nu_{\rm rel}$ (62) to the particle. The value
$\nu_{\rm rel}$ can be presented in the two equivalent forms:
$\nu_{\rm rel} = \sqrt{p^2c^2 + M_0^2c^4}/h =Mc^2/h$. This
frequency supposedly describes the wave behaviour of the particle.
In our model the hypothetical particle frequency $\nu_{\rm rel}$
is replaced by the frequency of collective vibrations
$\omega_{{\vec k}_{v_0}}$ of space singularity which surrounds the
particle. In accordance with expression (68) the values of the two
frequencies are the same:
  $$ \nu_{\rm rel}= \omega_{{\vec
k}_{v_0}}/2\pi = Mc^2/h. \eqno(73)
  $$
Hence a conclusion can be drawn that the Dirac wave equation (63)
represents exclusively the spectrum of the crystallite space
singularity that surrounds the particle rather than describes the
particle behaviour. Moreover such the interpretation of the origin
of the relativistic particle frequency makes possible to include
(due to the particle's inerton cloud) the information associated
with the particle spin into the spectrum $\omega_{{\vec
k}_{v_0}}/2\pi$ [see (59)]. The wave function of the Dirac
equation, i.e. the spinor, is partitioned to four components
$\psi_s$ ($s=1,\ 2,\ 3,\ 4$) which should set connections between
the system parameters inside the deformation coat surrounding the
particle. What is nature of spinor components? Evidently, in the
framework of the model that is considered herein the spinor
consists of two components:  one characterises proper vibrations
of the crystallite's superparticles and the other characterises
the inerton cloud that oscillates in the bounds of the crystallite
too. Each of these two components includes also the two possible
spin components, $\alpha = \downarrow,\uparrow$, which augment the
total number of spinor components to four.  Such structure is
imposed on the ${\hat{\vec \alpha}}$-matrix as well: the
${\hat{\vec\alpha}}$ contains information on the energy of the
crystallite, on the energy of the inerton cloud, and on the
energies of their spin components.

       Elastic vibrations of massive superparticles in the singularity
region  are nothing but standing gravitational waves, the
particle's inertons being virtual standing gravitational waves as
well. But when relating this region and its wave excitations to
gravitation, involved in the Dirac equation must be the terms
which might be interpreted as a manifestation of gravitation. It
is of interest that Chapman and Cerceau [51] arrived precisely at
this conclusion. In the above paper they managed to bring the
Dirac equation into the form from which it may be seen that the
particle spin immediately interacts with the background
gravitational field. To their opinion, this result calls for an
explanation.

\vspace{4mm}

\section{\bf Conclusion}

\vspace{2mm} \hspace*{\parindent}
      The present paper sets out to demonstrate the existence of the
hidden mechanism that makes it possible to construct the universal
quantum theory developed in real space which is capable to unite
the two limit cases -- nonrelativistic and relativistic quantum
theories. In the proposed approach the central role is played by
cellular elastic space: it forms particles and determines their
behaviour providing the particles with oscillating motion.  The
concept of cellular elastic space has also helped to solve the
spin problem, which has been reduced to special intrinsic particle
oscillations.

     It has been shown in the author's preceding papers [32-35] and in
the present work that the kinetics of a particle in space -- the
parameters $\lambda$ and $\Lambda$ can be considered as the free
path lengths for the particle and its inerton cloud respectively
-- can easily result in the Schr\"odinger and Dirac formalism,
with the satisfaction of all formulas of special relativity.
Besides for the first time the theory permits to investigate the
nature of the phase transition which takes place in a quantum
system when we turn from the description based on the
Schr\"odinger equation to that resting on the Dirac one. Moreover
the submicroscopic consideration of the particle behaviour in
space allows us to conclude that gravitons of general relativity
as carriers of gravitational interaction do not exist and they
should make way for inertons, space elementary excitations, or
quasi-particles, which always accompany any particle when it
moves.

Finally, we can infer as corollary that 1) the gravitational
radius of an absolutely rested particle is bounded by the size of
the space crystallite singularity enclosing the particle (i.e.,
the half of the Compton wavelength $h/2Mc$) and 2) the
gravitational radius of a moving particle is restricted by the
amplitude $\Lambda=\lambda c/v_0$ of inerton cloud oscillating in
the vicinity of the particle along the whole particle path.

 At the same time the research conducted raises a new
significant problem. It is necessary to prove that the space net
deformation, which is transferred by inertons, obeys the rule
$1/r$, i.e., that inertons indeed play a role of real carriers of
the gravitational interaction describing by the Newton law.

\vspace{8mm}

\section*{\large \bf Appendix}

\vspace{2mm} \hspace*{\parindent} Let us consider the space
crystal singularity incorporating $N$ superparticles. Assume that
$m_{\vec n}$ is the superparticle mass and \ $\zeta _{{\vec
n}{\beta}}$ \ ($\beta $= 1, 2, 3) are the three components of the
superparticle displacement from the centre of the cell defined by
the lattice vector ${\vec n}$. Superparticles interact only with
the nearest neighbors and consequently, if the position of a
certain cell is depends on the vector $\vec n$, its nearest
neighbor may be described by the vector \ ${\vec n}+ {\vec a}$ \
($a$ is the crystallite structure constant). The Lagrangian of the
lattice in question in the harmonic approximation appears as $$
L=\frac 1{2}\sum_{\vec n, \beta}m_{\vec n}\dot\zeta^2_{\vec n
\beta} -\frac 1{2} \sum_{\vec n,\beta\beta^{\prime}} \gamma_{\beta
\beta^{\prime}} (\zeta_{\vec n\beta}- \zeta_{\vec n-\vec a,
\beta})^2 \eqno(A1) $$ where $\gamma_ {\beta \beta^{\prime}}$ is
the space-crystal elasticity tensor.

       In the solid state theory, for the transition to the collective
variables $A_{\vec k}= (A_{-\vec k})^*$ in (A1) with $m_{\vec
n}=m={\rm const}$, use is made (see, e.g.  Davydov [52])  of the
canonical
 transformation
$$ \zeta_{{\vec n}\beta}=\frac 1{\sqrt{N}}\sum_{\vec k}e_{\beta}
(\vec k) A_{\vec k}\exp(i\vec k \vec n)
\eqno(A2)
$$
  where the quantities
$e_{\beta}(\vec k)$ represent the three vibration branches (one
longitudinal and two transverse). Subsequent transformations lead to
the Hamiltonian function
$$
H=\frac 12\sum_{\vec k,
\beta}\Bigl[\frac 1{m}P_{\vec k \beta}P_{-\vec k \beta} + m
\Omega^2_{\beta}(\vec k) A_{\vec k \beta}A_{-\vec k \beta} \Bigr]
\eqno(A3)
$$
and then to the Hamiltonian  operator
$$
\hat H =\frac 12 \sum_{\vec k,
\beta}\hbar\Omega_{\beta}(\vec k) \bigl[\hat b^+_{\vec k \beta}\hat
b_{\vec k \beta}+1/2 \bigr].  \eqno(A4)
$$
The energy $E$
of vibrations of the solid crystalline lattice is determined according
to the formula
$$
 E=\sum_{\vec k,
\beta}\hbar\Omega_{\beta}(\vec k)f_{\vec k}
\eqno(A5)
$$
where $f_{\vec k}$ is the Planck distribution function for
phonons. As is seem from (A5), the energy spectrum of a solid
crystal is the sum of a complete vibration set.

      However, in the present case of the space crystallite the
situation is different. When a particle is born, the crystallite
is formed adiabatically quickly around it, with the speed $c$
ultimate for degenerate space. Therefore, if stands to reason to presume
that when crystallite is formed, the superparticles involved
are coherently excited and, as a result, the whole crystallite,
being at a zero temperature, appears to be in  only one, the
lowest excited state. Hence the transition to the collective
variables in (A1) may not incorporate superposition of states,
that is the transformation to the variable $A_{\vec k}$ should be
selected in the form of
$$
\zeta_{\vec n \beta}=e_{\beta}(\vec k)A_{\vec k}\exp(i\vec k
\vec n_{\beta})
\eqno(A6)
$$
 which reflects the specific initial condition of the crystallite
formation. Substitution of (A6) into (A1) gives the space
crystallite Lagrangian  written in new coordinates and their
derivatives:
$$
  L=\frac 12 \sum_{\vec n}(m_{\vec n}\dot A_{\vec
k}\dot A_{\vec k}^* - m_{\vec n}\omega^2_{\vec k}A_{\vec k}A_{\vec
k}^*)
\eqno(A7)
$$
where the notation
$$
\sum_{\vec n}m_{\vec
n}\omega^2(\vec k)=\sum_{\vec n} \delta_{\vec n \vec n} \ 4\sum_{\beta
\beta^{\prime}}\gamma_ {\beta \beta^{\prime}}\sin^2(\vec k \vec
a_{\beta^{\prime}})
\eqno(A8)
$$
is introduced. In the long-wave approximation, $ak \ll 1$, instead of
(A8) one  derives
$$ \omega^2_{\vec k}\simeq N\sum_{\beta
\beta^{\prime}} \gamma_{\beta \beta^{\prime}}(\vec k \vec a_
{\beta^{\prime}})^2/m_{\Sigma} \eqno(A9) $$ where $m_{\Sigma}= \sum
_{\vec n}m_{\vec n}$.  Following from (A9) is the expression for the
velocity of the collective elastic vibration (standing wave) in the
crystallite $$ c=\omega_{\vec k}/k_N = a \Bigl( N\sum_{\beta
\beta^{\prime}}\gamma_ {\beta\beta^{\prime}}/m_{\Sigma} \Bigr)^{1/2}.
\eqno(A10)
$$

         From (A7) it follows that the generalized momentum
$$ P_{\vec k}=\partial L /\partial \dot A_{\vec k}=m_{\Sigma} \dot
A^*_{\vec k}. \eqno(A11) $$ By means of (A7) and (A11) we deduce
the expression for the space crystallite energy $$ E_{\vec
k}=|P_{\vec k}|^2/2m_{\Sigma}+m_{\Sigma}\omega^2_{\vec k} |A_{\vec
k}|^2/2. \eqno(A12) $$ We now enter the action function and,
having regard to (A12), pass on to the Hamilton-Jacobi equation $$
( \partial S_{\vec k}/ \partial |A_{\vec k}|)^2 / 2 m_{\Sigma}+
m_{\Sigma} \omega^2_{\vec k} |A_{\vec k}|^2 / 2 = E_{\vec k}.
\eqno(A13) $$ From (A13) in the action-angle variables (see ter
Haar [53]) for the action increment within the period we obtain $$
 J=E_{\vec k}/(\omega_{\vec k}/2\pi).
\eqno(A14)
$$

Assuming that $J/2\pi=\hbar$ from (A14) we have
$$
         E_{\vec k}= \hbar \omega_{\vec k}.
\eqno(A15)
$$

\vspace{8mm}

\section*{\large \bf Acknowledgement}

\vspace{2mm}

\hspace*{\parindent} I am very grateful to Prof. M.~Bounias for
the discussion of the work and thankful to Prof. F.~Winterberg and
Dr. X.~Oudet placed at my disposal their works quoted in the
present paper. Many thanks also to Prof. J.~Gruber provided me
with works of Dr. H.~Aspdent cited in this paper as well.


\newpage
\begin{thebibliography}{99}
\bibitem{1} S. A. Goudsmit,  It might as well as spin,
           {\it Phys.  Today} {\bf 29}, no. 6, 40-43 (1976).

\bibitem{2} G. E. Uhlenbeck,  Personal  reminiscences, {\it Phys.
            Today} {\bf 29}, no. 6,  43-48 (1976).

\bibitem{3} B. L. van der Waerden, Pauli  exclusion principle
and spin, in: {\it Theoretical physics in the twentieth century},
eds.: M.  Fiersz and V. F. Weisskopf (Izdatelstvo inostrannoy
literatury, Moscow, 1962) pp.  231-284 (Russian translation).

\bibitem{4} E. Schr\"odinger,  \"Uber die kraftefreie Bewegung in der
relativistischen Quantenmechanik, {\it Sitz. Preuss. Acad.
Wissen., Phys.-Math. Kl.} {\bf XXIV}, SS. 418-428  (1930).

\bibitem{5} P. A. M. Dirac,  The  principles of quantum
mechanics (Nauka, Moscow, 1979), p. 345  (Russian translation).

\bibitem{6} Ya. I. Frenkel, {\it Electrodynamics}, vol. 1 (Gosudarstvennoe
tekhniko-teoreticheskoe izdatelstvo,  Leningrad -- Moscow, 1934),
p. 371 (in Russian).

\bibitem{7} F. Halbwachs, J. M. Souriau, and J.-P. Vigier,  Le
groupe d'invariance associ\'e  aux rotateurs relativistes et la
th\'eorie bilocale, {\it J. de Phys. et le Radium} {\bf 22},
393-406 (1961).

\bibitem{8} V. L. Ginzburg and V. I. Man'ko,  Relativistic
oscillator models of elementary particles, {\it Nucl. Phys.}, {\bf
74}, 577-588 (1965).

\bibitem{9} N. C. Petroni, Z. Maric, Dj. Zivanovich, and J.-P. Vigier,
Stable  states of a relativistic bilocal stochastic oscillator:  a
new quark-lepton model, {\it J. Phys. A: Math. Gen.}, {\bf 14},
501-508 (1981).

\bibitem{10} E. Plahte, Interrelationships  of quantum and
classical spin-particle theories, {\it Suppl. Nuovo Cimen.}, {\bf
5}, 944-953 (1967).

\bibitem{11} M. Umezawa,  Trembling  motion of the free spin 1/2
particle, {\it Progr. Theor. Phys.} {\bf 71}, 201-208 (1984).

\bibitem{12} A. O. Barut and W. D. Tracker,  Zitterbewegung of
the electron in external fields, {\it Phys. Review D} {\bf 31},
2076-2088 (1985).

\bibitem{13} A. Bohm,  L. J. Boya, P. Kielanowski, M. Kmiecik,
M. Loewe, and P. Magnollay, Theory  of relativistic extended
objects, {\it Int. J. of Modern Phys.} {\bf 3}, 1103-1121 (1988).

\bibitem{14} A. O. Barut and N. Zanghi, Classical  model of the
Dirac electron, {\it Phys. Rev. Lett.} {\bf 52},  2009-2012
(1984).

\bibitem{15} G. Spavieri,  Model  of the electron spin in stochastic
physics, {\it Foud. Phys.} {\bf 20}, 45-61 (1990).

\bibitem{16} F. A. Berezin and M. S. Marinov,  Particle  spin
dynamics as a Grassmann variant of classical mechanics, {\it Ann.
Physics} {\bf 104}, 336-362 (1977).

\bibitem{17} P. P. Srivastava  and N. A. Lemos, Supersymmetry
and classical particle spin dynamics, {\it Phys. Rev. D} {\bf 15},
3568-3574 (1977).

\bibitem{18} V. V. Kuryshkin and E. E. Entralgo,  On  classical
theory of a point particle with spin, {\it Dokl. Akad. Nauk USSR}
{\bf 312},  350-353 (1990) (in Russian);   Structural-point
objects and classical model of spin, {\it ibid.} {\bf 312},
592-596 (1990) (in Russian).

\bibitem{19} H. C. Ohanian,  What  is spin?, {\it Am. J. Phys.} {\bf 54},
500-505 (1986).

\bibitem{20} A. Heslot, Classical  mechanics and the electron spin,
{\it Am. J. Phys.} {\bf 51}, 1096-1102 (1983).

\bibitem{21} K. Shima,  On  the spin-1/2 gauge field, {\it Phys.
Lett. B} {\bf 276}, 462-464 (1992).

\bibitem{22} M. Pav$\rm \check s$i$\rm \check c$, E. Recami,
W. A. Rodriges (Jr), G. D. Maccarrone, F. Raciti, and G. Salesi,
Spin and electron structure, {\it Phys. Lett. B} {\bf 318},
481-488 (1993).

\bibitem{23} H. C. Corben,  Structure  of a spinning point
particle at rest, {\it Int. J. Theor. Phys.} {\bf 34}, 19-29
(1995).

\bibitem{24} B. G. Sidharth, The symmetry underlying spin and the
Dirac equation: footprints of quantized space-time,  arXiv.org
e-print archive \ quant-ph/98111032.

\bibitem{25} X. Oudet, L'aspect corpusculaire des \'electrons et la
notion de valence dans les oxydes m\'etalliques, {\it Ann. de la
Fond. L.  de Broglie} {\bf 17}, 315-345 (1992);   L'\'etat
quantique et les notions de spin, de fonction d'onde et d'action,
{\it ibid.} {\bf 20},  473-490 (1995);  Atomic magnetic moments
and spin notion, {\it J. App.  Phys.} {\bf 79},  5416-5418 (1996).

\bibitem{26} D. Bohm, On  the role of hidden variables in the
fundamental structure of physics, {\it Found. Phys.} {\bf 26},
719-786 (1996).

\bibitem{27} P. I. Fomin, Zero  cosmological constant and Planck
scales phenomenology, in: {\it Quantum gravity. Proceedings of
the fourth seminar on quantum gravity}, Moscow, USSR, 1987, eds.:
V. Markov, V.  Berezin, and V. P. Frolov (World Scientific
publishing Co., Singapore, 1988), pp.  813-823.

\bibitem{28} H. Aspden, The  theory of the gravitational constant,
{\it Phys. Essays} {\bf 2}, 173 - 179 (1989);  The  theory of
antigravity, {\it Phys. Essays} {\bf 4}, 13-19 (1991);  {\it
Aetherth Science Papers} (Subberton Publications,  P. O. Box 35,
Southampton SO16 7RB, England, 1996).

\bibitem{29} J. W. Vegt,  A  particle-free model of matter based
on electromagnetic self-confinement (III), {\it Ann. de la Fond.
L. de Broglie} {\bf 21}, 481-506 (1996).

\bibitem{30} F. Winterberg,  Physical  continuum and the problem of
a finistic quantum field theory, {\it Int. J. Theor. Phys.} {\bf
32}, 261-277 (1993);  Hierarchical order of Galilei and Lorentz
invariance in the structure of matter, {\it ibid.} {\bf 32},
1549-1561 (1993);  Equivalence and gauge in the Planck-scale
aether model, {\it ibid.} {\bf 34}, 265-285 (1995);
Planck-mass-rotons cold matter hypothesis, {\it ibid.} {\bf 34},
399-409 (1995); Derivation of quantum mechanics from the Boltzmann
equation for the Planck aether, {\it ibid.} {\bf 34}, 2145-2164
(1995); Statistical mechanical interpretation of hole entropy,
{\it Z. Naturforsch.} {\bf 49a}, 1023-1030 (1994); Quantum
mechanics derived from Boltzmann's equation for the Planck aether,
{\it ibid.}, {\bf 50a},  601-605 (1995).

\bibitem{31} A. Rothwarf, An aether model of the universe, {\it Phys.
Essays} {\bf 11}, 444-466 (1998).

\bibitem{32} V. Krasnoholovets and D. Ivanovsky, Motion  of a
particle and the vacuum, {\it Phys. Essays} {\bf 6}, 554-563
(1993) (also arXiv.org e-print archive \ quant-ph/9910023).

\bibitem{33} V. Krasnoholovets,  Motion of a relativistic particle
and the vacuum, {\it Phys. Essays} {\bf 10}, 407-416 (1997) (also
arXiv.org e-print archive \  quant-ph/9903077).

\bibitem{34} V. Krasnoholovets, On the way to submicroscopic
description of nature, ( also arXiv.org e-print archive \
quant-ph/9908042.

\bibitem{35} P. G. Bergmann, {\it Introduction  to the theory of
relativity} (Gosudarstvennoe izdatelstvo inostrannoy literatury,
Moscow, 1947), p. 205 (Russian translation).

\bibitem{36} See Ref. 36, p. 235.

\bibitem{37} W. Pauli, {\it Theory of relativity} (Nauka, Moscow,
1983), p. 205 (Russian translation).

\bibitem{38} S. Weinberg, {\it Gravitation  and cosmology: principles
and applications of the general theory of relativity} (Mir,
Moscow, 1975), p.  138 (Russian translation).

\bibitem{39} B. A. Dubrovin,  S. P. Novikov, and A. T. Fomenko,
{\it Modern  geometry: methods and applications} (Nauka, Moscow,
1986), p.  372 (in Russian).

\bibitem{40} A. Ashtekar, {\it Lectures on non-perturbative canonical
gravity} (World Scientific, Singapore, 1991).

\bibitem{41} B. G. Sidharth, Quantum mechanical black holes: towards
a unification of quantum mechanics and general relativity, {\it
Ind. J. Pure Appl. Phys.} {\bf 35}, 456-471 (1997) (also arXiv.org
e-print archive \ quant-ph/9808020).

\bibitem{42} L. M. Krauss, Cosmological antigravity, {\it Sc. Amer.}
            {\bf 280}, no. 1, 52-59 (1999).

\bibitem{43} V. Krasnoholovets, On the theory of the anomalous
      photoelectric effect stemming from a substructure of matter waves,
      {\it Ind. J. Theor. Phys.}, in press (also arXiv.org e-print archive
      \ quant-ph/9906091).

\bibitem{44} D. ter Haar,  {\it Elements of Hamiltonian mechanics}
     (Nauka, Moscow, 1974), p. 157 (Russian translation).

\bibitem{45} H. Goldstein, {\it Classical  mechanics} (Nauka, Moscow,
       1974), p. 306 (Russian translation).

\bibitem{46} L. de Broglie, {\it Heisenberg's uncertainty relations and
       the probabilistic interpretation of wave mechanics} (Mir, Moscow,
        1986), p.  42 (Russian translation).

\bibitem{47} S. Sonego, Conceptual  foundations of quantum theory:
     a map of the land, {\it Ann. de la Fond. L. de Broglie} {\bf 17},
     405-473 (1992); errata: {\it ibid.} {\bf 18},  131-132 (1993).

\bibitem{48} A. A. Sokolov, Yu. Loskutov,  and L. M. Ternov,  {\it
      Quantum Mechanics} (Prosveshchenie, Moscow, 1965), p. 306 (in
         Russian).

\bibitem{49} V. L. Ginzburg,  {\it About  physics and astrophysics},
          (Nauka, Moscow, 1985), p. 124 (in Russian).

\bibitem{50} L. de Broglie,  Interpretation  of quantum mechanics
         by the double solution theory, {\it Ann. de la Fond. L. de
         Broglie} {\bf 12},  399-421 (1987).

\bibitem{51} T. C. Chapman and O. Cerceau,  On  the
        Pauli-Schr\"odinger equation, {\it Am. J. Phys.} {\bf 52}, 994-997
        (1984).

\bibitem{52} A. S. Davydov, {\it  The theory of solids} (Nauka, Moscow,
             1976), p. 46 (in Russian).

\bibitem{53} See Ref. [44], p. 165.

\end{thebibliography}
\end{document}